\documentclass[sigconf,nonacm]{acmart}

\def\DiffMin/{\textsc{DiffMin}}

\usepackage{enumitem}
\usepackage[linesnumbered, algoruled, lined,ruled]{algorithm2e}

\title{Refining Fuzzed Crashing Inputs for Better Fault Diagnosis}

\author{Kieun Kim}
\email{kieun@cbnu.ac.kr}
\affiliation{%
  \institution{Chungbuk National University (CBNU)}
  \city{Cheongju}
  \country{Republic of Korea}
}

\author{Seongmin Lee}
\email{seongmin.lee@mpi-sp.org}
\affiliation{%
  \institution{Max Planck Institute for Security and Privacy (MPI-SP)}
  \city{Bochum}
  \country{Germany}
}

\author{Shin Hong}
\email{hongshin@cbnu.ac.kr}
\affiliation{%
  \institution{Chungbuk National University (CBNU)}
  \city{Cheongju}
  \country{Republic of Korea}
}
\acmConference{}{}{}
\begin{document}

\begin{abstract}
We present \DiffMin/, a technique that refines a fuzzed crashing input to gain greater similarities to given passing inputs to help developers analyze the crashing input to identify the failure-inducing condition and locate buggy code for debugging.
\DiffMin/ iteratively applies edit actions to transform a fuzzed input while preserving the crash behavior. Our pilot study with the Magma benchmark demonstrates that \DiffMin/ effectively minimizes the differences between crashing and passing inputs while enhancing the accuracy of spectrum-based fault localization, highlighting its potential as a valuable pre-debugging step after greybox fuzzing.
\end{abstract}


\settopmatter{printacmref=false}
\setcopyright{none}

\maketitle
\pagestyle{empty}
\section{Introduction}
\label{sec:intro}

Greybox fuzzers~\cite{manes2019art} generate new test inputs by repeatedly applying random mutations to existing test inputs while guiding the random mutation process to maximize code coverage. 
Tools such as AFL++~\cite{fioraldi2020afl++} and libFuzzer~\cite{serebryany2016continuous} have proven highly effective in generating crashing inputs, and are now widely used for testing open-source projects~\cite{serebryany2017oss} to assist project maintainers to detect and mitigate reliability and security threats quickly.

When a fuzzer generates a crashing input, the project maintainers diagnose the failure for debugging. 
Bisection aids fault diagnosis by identifying bug-inducing commits if the fuzzing infrastructure supports conducing fuzzing across program versions~\cite{abreu2021reducing}. Otherwise, the maintainers need to inspect the input data to determine which aspects of the input trigger the failure. 
In addition, they analyze the code coverage or execution traces produced by the input to locate code elements that contribute significantly to the crash~\cite{wong2016survey, jones2005empirical}. 

A key challenge in fault diagnosis is that fuzzed crashing inputs are often difficult for human maintainers to analyze, since these fuzzed inputs typically differ significantly from valid program inputs due to the accumulation of random mutations. Moreover, since greybox fuzzers generate these inputs while aiming to maximize code coverage, fuzzed crashing inputs typically explore diverse program features, many of which are unrelated to the crashes.

To aid fault diagnosis after fuzzing, we present \DiffMin/, a technique that refines a fuzzed crashing input into a crashing input with greater similarity to a given passing input.
Given a pair of crashing and passing inputs, \DiffMin/ first compares the two inputs and identifies edit actions to convert the crashing input to the passing input. Subsequently, \DiffMin/ iteratively applies each edit to the crashing input until the crash disappears. As a result, \DiffMin/ derives a refined crashing input that shares more aspects with the passing input, than the original fuzzed crashing inputs. We suspect that crashing inputs refined by \DiffMin/ would be better to diagnose, compared to the original fuzzed crashing inputs.

We have conducted a pilot study to demonstrate the efficacy of \DiffMin/ with the Magma benchmark~\cite{hazimeh2020magma}. From the study results with four buggy programs, we found that \DiffMin/ effectively transforms fuzzed crashing inputs into alternative crashing inputs having greater similarities with given passing inputs, reducing the differences between crashing and passing inputs. In addition, we found, for three programs out of the four, that the accuracies of spectrum-based fault localization are improved when refined inputs are used, instead of the original fuzzed crashing inputs.

\section{\DiffMin/: Fuzzed Crashing Input Refinement Technique}
\label{sec:tech}


\begin{algorithm}[t]
        \small
        \SetKwFunction{GetEdits}{GetEdits}
        \SetKwRepeat{Do}{do}{while}

        \caption{\DiffMin/}
        \label{diffmin} 
                
\KwIn{
    $P$, a program under test; $c$, a crashing input to refine; $p$, a passing input 
}
      
\KwOut{
    $c_{min}$, a refined crashing input
}

\BlankLine        
$c_{min} \gets c$ ;\\
\Do{$c' \neq \bot$}{
    $Edits \gets GetEdits(p,c_{min})$ ; \\
    $c' \gets \bot $ ; \\
    \ForEach{$e \in Edits$} {
        $c_{e}  \gets EditApply(c, e)$ ; \\
        \If{$P(c) = P(c_{e})$ /* crashing preserved */}
        {
            \If{$c' = \bot \vee 
                EditDist(p, c_{e}) < EditDist(p, c')$} {
               $c' \gets c_{e}$ ;
            }
        }    
    }
    \If{$c' \neq \bot$} {
        $c_{min} \gets c'$ ;
    }
    
}
\Return{$c_{min}$}

\end{algorithm}


We present \DiffMin/, a new technique for refining crashing inputs to enhance fault diagnosis, 
as shown in Algorithm~\ref{diffmin}. \DiffMin/ takes as input a program under test ($P$), and 
both a crashing input ($c$) and a passing input ($p$): 
the crashing input would be a fuzzed crashing input that may be difficult for project maintainers 
to analyze, and the passing input is a valid input that they are familiar with. 
The given passing input ($p$) serves as a reference to which the crashing input ($c$) must be 
transformed to achieve similarity.
Using these two test inputs ($p$ and $c$), \DiffMin/ generates a new crashing input ($c_{min}$) 
with a reduced lexical distance from the passing input compared to the original crashing input. 

The key idea behind \DiffMin/ is to define a series of edits by computing the differences between 
the two given inputs and iteratively applying each edit while preserving the same crashing behavior,
as inspired by delta debugging~\cite{zeller2002simplifying}. Note that \DiffMin/ is different from
delta debugging because \DiffMin/ focuses on minimizing the difference between a crashing input and 
a passing input rather than merely reducing the size of the crashing input.

Starting with the given crashing input (Line 1), \DiffMin/ iteratively refines the crashing input by 
applying one edit that minimizes the lexical difference and reproduces the same crash (Lines 2-16).
At each iteration, \DiffMin/ represents 
both inputs as byte strings and uses the Hirshberg's algorithm~\cite{hirschberg1975linear} to 
find an optimal sequence alignment between the two byte strings which minimizes their Levenshtein 
distance. Given sequence alignment, 
\DiffMin/ derives a set of possible edits ($Edits$) by defining each substring insertion, 
deletion or replacement as an edit operation (Line 3).
Once possible edits are defined, \DiffMin/ iterates over these edits (Lines 5-12) to apply each edit (Line 6)
and finds one that reduces the Levenshtein distance (Line 8) most while reproducing the same crash (Line 7).
If such edit exists ($c'$), \DiffMin/ updates the latest minimized crashing input (Lines 13-14) 
and takes another iteration (Line 16).
If there exists no single edit preserving the same crash behavior, \DiffMin/ returns the
latest minimized crashing input as output (Line 17). 



\section{Pilot Study}

We conducted a pilot study to evaluate whether \DiffMin/ effectively refines fuzzed crashing inputs and assists in fault diagnosis tasks. Specifically, we designed this study to answer two questions: (1) To what extent does \DiffMin/ reduce the lexical distance between a fuzzed crashing input and a passing input (RQ1)? and (2) Does SBFL results improve when refined crashing inputs are used instead of fuzzed crashing inputs (RQ2)?
We selected four buggy programs from Magma~\cite{hazimeh2020magma} as target programs (Table~\ref{tbl:bench}). These four programs were randomly sampled from the 138 buggy programs available in Magma. Since each Magma target program contains multiple bugs, we configured each program to include only the target bug by disabling the other bugs and enabling the bug-specific test oracle (i.e., canary). We obtained one crashing input by running AFL++, which took between 4 and 230 minutes.

{\small
\begin{table}[t]
\vspace{-0.15in}
    \centering
    \caption{Target Buggy Programs from Magma~\cite{hazimeh2020magma}}
    \label{tbl:bench}    
    \vspace{-0.1in}
    \begin{tabular}{l r r r r}
        \toprule
        Bug ID & 
        Num. &
        Avg. Initial&
        Time to Crash &
        Crashing  \\
         & 
        Initial Seeds &
        Seed Size &
         &
        Input Size \\
        \midrule

    PNG006 &	4	& 312 bytes	& 4	min & 384 bytes \\
    PNG007 &	4	& 312 bytes	& 230 min	& 281 bytes \\
    XML003 &	1196 &	505	bytes & 36 min	& 33024 bytes \\
    XML009 &	1196	& 505 bytes	& 104 min	& 233 bytes \\

\bottomrule
   
    \end{tabular}
      
    \end{table}
}

To answer RQ 1, we applied \DiffMin/ for each initial seeds (i.e., passing inputs) and compared the Levenshtein distances between the initial seeds and the fuzzed crashing inputs (labeled as $Dist(p,c)$), as well as the distances between the initial seeds and the \DiffMin/ results (labeled as $Dist(p, $\DiffMin/$c_{min})$). Table~\ref{tbl:rq1} shows the minimal, average, and maximal Levenshtein distances (in bytes) with all initial seeds. These results clearly show that \DiffMin/ significantly reduces the lexical distances in most cases. For the four programs, the ratios of the average distances with the \DiffMin/ results to the average distances with the fuzzed crashing inputs are 14\%, 90\%, 45\% and 40\%, respectively. For example of XML003, \DiffMin/ transforms the fuzzed crashing input having 33024 bytes into a 174-bytes crashing input subject to a 90-bytes passing input.

To answer RQ 2, we conducted SBFL with three set-ups: (1) use all initial seeds (passing tests) and the fuzzed crashing input (labeled as {\it fuzz} in Table~\ref{tbl:rq2}), (2) use all initial seeds and the ddmin~\cite{zeller2002simplifying} result of the fuzzed crashing input ({\it ddmin}), (c) use all initial seeds and all crashing inputs refined by \DiffMin/ with the initial seeds (\DiffMin/). Table~\ref{tbl:rq2} shows the best ranks of a buggy line (statement-level) and a buggy function with Op2. The result shows that \DiffMin/ improves statement-level rankings for three programs, and function-level rankings for two programs, suggesting that \DiffMin/ can substantially contributes to improving SBFL accuracy.

{\small
\begin{table}[t]
    \centering
    \caption{RQ 1. Lexical Distance Reduction}
    \vspace{-0.15in}
    \begin{tabular}{l | r r r |  r r r}
        \toprule
     
        &\multicolumn{3}{c|}{$Dist(p,c)$}
        &\multicolumn{3}{c}{$Dist(p,$\DiffMin/$(c,p))$} \\
        & Min & Avg & Max & 
        Min & Avg & Max \\
        \midrule

    PNG006 &	
    341 &	348	& 356	& 5	& 52	& 123 \\
    PNG007 &	
    151	& 240	& 320 &	137 &	217 & 	285 \\
    XML003 &	
    72	& 32681 &	45533 &	72 &	14777 & 	33023 \\
    XML009 &	
    7 & 1663 & 140462 & 4 & 667 & 106468 \\

\bottomrule
   
    \end{tabular}
    \label{tbl:rq1}    
    \end{table}
}

{\small
\begin{table}[t]

    \centering
    \caption{RQ 2. SBFL Result (Op2)}
    \label{tbl:rq2}    
    \begin{tabular}{l | r r r |  r r r}
        \toprule
     
        &\multicolumn{3}{c|}{Statement-level}
        &\multicolumn{3}{c}{Function-level} \\
        & {\it fuzz} & {\it ddmin} & \DiffMin/  
        & {\it fuzz} & {\it ddmin} & \DiffMin/ \\
        \midrule

    PNG006 &	
    70 &	70	& {\bf 50}	& 16	& 16	& {\bf 14} \\
    PNG007 &	
    70	& 70	& {\bf 58} &	{\bf 12} &	{\bf 12} & {\bf 12} \\
    XML003 &	
    116	& 89 &	{\bf 29} &	14 & 12 & {\bf 4} \\
    XML009 &	
    95 & {\bf 94} & {\bf 94} & {\bf 12} & {\bf 12} & {\bf 12} \\

\bottomrule
   
    \end{tabular}
    \end{table}
}

\section{Limitations and On-going Works}

 Our pilot study using four buggy programs from the Magma benchmark demonstrates that \DiffMin/ effectively minimizes the differences between crashing and passing inputs while enhancing the accuracy of spectrum-based fault localization, highlighting its potential as a valuable pre-debugging step after greybox fuzzing.
Although the pilot study shows positive results, the proposed technique and empirical evaluations still have certain limitations. The following discussion explores these limitations and the ongoing work addressing them from different perspectives:
\begin{itemize}[leftmargin=0.35cm]
\item {\bf Crash Input Refinement Algorithm.} The crashing inputs that \DiffMin/ can discover are limited by the possible edit operations defined through lexical comparison. For example, if an edit containing the failure-inducing aspect of the input is defined too coarsely, \DiffMin/ may lose the opportunity to further reduction. We will explore alternative refinement algorithms that define and apply edits using different strategies. 

\item {\bf Fault Diagnosis Difficulty.} Currenlty, we used lexical distances as the complexity measure. However, this is limited to our assumption that maintainers inspect a crashing input by differencing it with initial seeds. We need to explore different metrics considering different fault diagnosis methods.  
\item {\bf SBFL Study.} We will conduct comprehensive empirical evaluations on how crashing input refinements influence the accuracies of spectrum-based fault localization. We will explore different strategies for selecting fuzzing inputs and generating crashing inputs to better understand the effectiveness of the proposed technique.
\end{itemize}

\bibliographystyle{ACM-Reference-Format}
\bibliography{references}

\end{document}